\documentclass{article}
\usepackage{dcase2023,amsmath,graphicx,url,times,booktabs, tabularx, hhline}

\title{Frequency \& Channel Attention for \\ Computationally Efficient Sound Event Detection}

\name{Hyeonuk Nam, Seong-Hu Kim, Deokki Min, Yong-Hwa Park \thanks{This work was supported by the Institute of Civil Military Technology Cooperation funded by the Defense Acquisition Program Administration and Ministry of Trade, Industry and Energy of Korean government under grant No. UM22409RD4, and Korea Research Institute of Ships and Ocean engineering a grant from Endowment Project of “Development of Open Platform Technologies for Smart Maritime Safety and Industries” funded by Ministry of Oceans and Fisheries(PES4880).}}
\address{Korea Advanced Institute of Science and Technology, South Korea\\
\{frednam, seonghu.kim, minducky, yhpark\}@kaist.ac.kr}

\begin{document}

\ninept
\maketitle

\begin{sloppy}

\begin{abstract}
We explore on various attention methods on frequency and channel dimensions for sound event detection (SED) in order to enhance performance with minimal increase in computational cost while leveraging domain knowledge to address the frequency dimension of audio data. We have introduced frequency dynamic convolution (FDY conv) in a previous work to release the translational equivariance issue associated with 2D convolution on the frequency dimension of 2D audio data. Although this approach demonstrated state-of-the-art SED performance, it resulted in a model with 150\% more trainable parameters. To achieve comparable SED performance with computationally efficient methods for practicality, we explore on lighter alternative attention methods. In addition, we focus on attention methods applied to frequency and channel dimensions. Joint application Squeeze-and-excitation (SE) module and time-frame frequency-wise SE (tfwSE) to apply attention on both frequency and channel dimensions shows comparable performance to SED model with FDY conv with only 2.7\% more trainable parameters compared to the baseline model. In addition, we performed class-wise comparison of various attention methods to further discuss various attention methods' characteristics.
\end{abstract}


\section{Introduction}
\label{sec:intro}
Sound event detection (SED), which aims to recognize a target sound event class and corresponding time localization within a given audio clip, has potential to be applied in various applications such as automation, robotics and monitoring \cite{CASSE, crnn, dcasebaseline}. In order to recognize and locate sound events, we need strong pattern recognition tools. Recent advances in deep learning (DL) methods brought significant progress in SED \cite{crnn, dcasebaseline}. While most works directly applied DL methods from other domains to SED without modification, few works adapted DL methods to SED by thoroughly analysing unique characteristics of audio data and sound events.

Frequency dimension has to be carefully considered when applying DL methods on audio-related DL applications. It is shown by previous works that methods considering frequency dimension significantly improved SED performance \cite{FDY, filtaug, mdfdy}. SED has been heavily relying on convolutional recurrent neural networks (CRNN) based architectures \cite{crnn, dcasebaseline}. 2D convolution in CRNN assumes shift-invariance on both time and frequency dimensions thus enforces translational eqauivariance on both dimensions \cite{FDY}. However, frequency is a shift-variant dimension where the same pattern sounds different when translated along the frequency dimension. At the same time, frequency exhibits loose shift-invariance within short frequency range thus slight pitch-shift does not harm auditory perception much. Thus frequency dimension is a delicate yet essential component to be considered for audio domain. 

In a previous study, we introduced frequency dynamic convolution (FDY conv) to release translational equivariance by 2D convolution on the frequency dimension of 2D audio data to consider its shift-variant characteristic \cite{FDY}. While FDY conv showed impressive performance on SED, it added 150\% more parameters to the model. However, in order to apply SED on various real applications, we might need to implement SED on devices with limited specifications. Thus, there is a need for computationally efficient SED methods which is lighter but as competent as current state-of-the-art models. To address this limitation and improve the practicality of SED models, we explore various lighter attention methods to enhance SED performance more efficiently. We aim to achieve this by addressing the frequency and channel dimensions, since those are two emphasized dimensions in audio domain \cite{FDY, mdfdy, c2datt}. Thus we experiment with various attention methods on frequency and channel dimensions. The main contributions of this work are:
\begin{enumerate}
    \itemsep0.3em
    \item{We explore various alternative attention methods which are computationally efficient for practicality, while considering channel and frequency dimensions to consider unique characteristics of audio domain.}
    \item{Joint application of squeeze-and-excitation (SE) and proposed time-frame frequency-wise SE (tfwSE) to re-weight both channel and frequency dimensions shows comparable performance to state-of-the-art method while only adding model parameters by 2.7\%.}
    \item{We discuss the characteristics of various attention methods on SED to provide further insights for practical implementation.}
\end{enumerate}
The official implementation code is available on GitHub\footnote{https://github.com/frednam93/lightSED}.

\section{Methods}
\label{sec:methods}
While frequency dynamic convolution (FDY conv) showed state-of-the-art performance and have been widely adopted on SED \cite{mdfdy, dcase2022_2nd, dcase2022_4th1, dcase2022_4th2,ATSED,pretrainedSED}, it adds considerable number of trainable parameters to the networks due to multiple basis kernels \cite{FDY, dyconv}. Thus there remains a need for sufficiently well-performing model with fewer parameters for practical applications. Since FDY conv's strength comes from attention mechanism which selectively focus on important elements of the input, we explore other alternative attention methods to achieve comparable performance.

\subsection{Variants of Squeeze-and-Excitation}

\begin{figure}[t]
\centerline{\includegraphics[width=8.55cm]{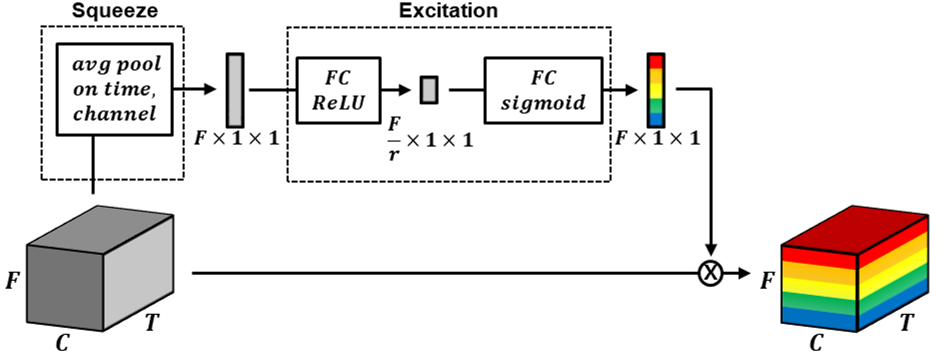}}
\caption{An illustration of frequency-wise Squeeze-Excitation.}
\label{fig:fwse}
\end{figure}

One alternative computationally efficient attention method widely used is squeeze-and-excitation (SE) \cite{senet}. It has been widely applied to various CNN-based models for its light yet powerful performance. SE module is composed of squeeze operation and excitation operation. Squeeze operation averages output of 2D convolution on two dimensions except channel to obtain squeezed intermediate representation. Excitation operation applies two successive fully connected (FC) layers to obtain attention weights representing relative importance of each channel. The channels of convolution output is re-weighted by multiplying the attention weight \cite{senet}. When applied to 2D audio data, squeeze operation is applied to the convolution output by:
\begin{equation}
z_c = \frac{1}{F \times T}\sum_{f=1}^{F}\sum_{t=1}^{T}x_{cft}
\end{equation}
\noindent where $z_c$ is intermediate representation after squeeze operation on $c$-th channel and $x_{cft}$ is the output by preceding 2D convolution with channel index $c$, frequency index $f$ and time index $t$. $F$ and $T$ are frequency and time dimension sizes of 2D convolution output. The excitation operation is composed of two FC layers as follows:
\begin{equation}
\label{excitation}
\mathbf{s} = \sigma (\mathbf{W_2} \delta (\mathbf{W_1 z}))
\end{equation}
\noindent where $\mathbf{s}$ is attention weight, also known as scale, which is multiplied to the output of preceding convolution. $\mathbf{z}$ is the intermediate representation vector. Both span channel dimension of size $C$. $\mathbf{W_1}$ and $\mathbf{W_2}$ are FC layers, $\delta$ refers to ReLU activation and $\sigma$ refers to sigmoid function.

To apply attention-based re-weighting on frequency dimension, Thienpondt \textit{et al.} proposed frequency-wise Squeeze-Excitation (fwSE) which applies SE on frequency dimension instead \cite{freqse}. Thus, instead of pooling time and frequency dimensions, fwSE pools channel and time dimensions during squeeze operation as follows:
\begin{equation}
z_f = \frac{1}{C \times T}\sum_{c=1}^{C}\sum_{t=1}^{T}x_{cft}
\end{equation}
\noindent The following excitation operation is the same as (\ref{excitation}), just that two FC layers are applied on frequency dimensions instead. Then, obtained attention weight for each frequency bin is multiplied to corresponding frequency components of the preceding convolution output. Fig. \ref{fig:fwse} illustrates the fwSE mechanism.

Since frequency component varies over time, we propose time-frame fwSE (tfwSE) which applies fwSE on every time frames of input instead of time-averaged input. Thus, tfwSE only pools channel dimension in squeeze operation and then applies excitation operation on every time frames. The squeeze operation on time frame $t$ can be expressed by following equation:
\begin{equation}
z_{ft} = \frac{1}{C}\sum_{c=1}^{C}x_{cft}
\end{equation}
\noindent where $z_{ft}$ is intermediate representation after squeeze operation. Then excitation is applied on frequency dimension on each time frame as follows:
\begin{equation}
\label{tfwexcitation}
\mathbf{s_t} = \sigma (\mathbf{W_2} \delta (\mathbf{W_1 z_t}))
\end{equation}
\noindent where $\mathbf{s_t}$ is scale on time frame $t$ and $\mathbf{z_t}$ is the intermediate representation vector corresponding to time frame $t$, both spanning channel dimension. Opposed to fwSE by Thienpondt \textit{et al.} which applies frequency-wise attention weights evenly over time-dimension by referring to representative averaged information of the convolution output, proposed tfwSE applies frequency-wise attention weights for each time frame by referring to each individual contents within corresponding time frame \cite{freqse}. While this could increase computation of excitation operator (fwSE applies excitation on one $\mathbf{z}$ per audio clip, while tfwSE applies excitation on $T$ $\mathbf{z_t}$ per audio clip), it could help generalizing excitation operation on many time frames. The mechanism of tfwSE is illustrated in Fig. \ref{fig:tfwse}. This method was previously applied by our submission on detection and classification of acoustic scenes and events (DCASE) 2022 challenge task 3 as well, showing its performance \cite{2022t3report}. Similarly, we could apply original SE on each time frame as well. We named it as time-frame SE (tSE). Note that this is not a time-wise version of SE like fwSE, as we do not apply SE by pooling channel and frequency dimensions during squeeze to leave time dimension. Instead, we pool frequency dimension only and apply SE on every time frame in similar way shown in Fig. \ref{fig:tfwse}.

\begin{figure}[t]
\centerline{\includegraphics[width=8.55cm]{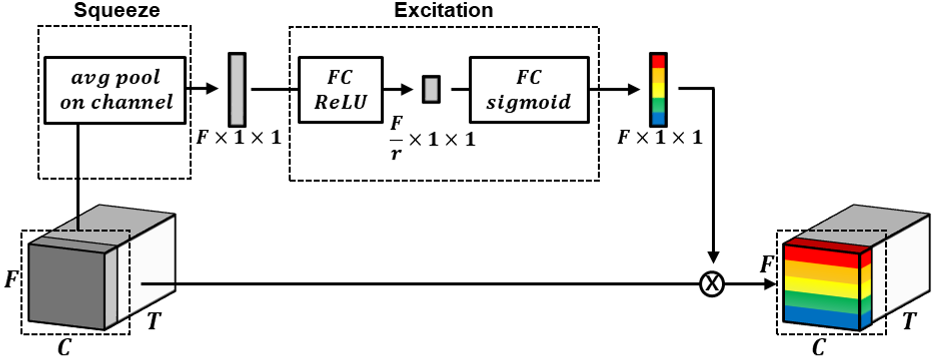}}
\caption{An illustration of time-frame frequency-wise Squeeze-Excitation on one time frame. tfwSE applies this procedure for every time frames.}
\label{fig:tfwse}
\end{figure}

\subsection{Channel-Frequency Attention Methods}
\label{C2D-Att}
Li \textit{et al.} \cite{c2datt} proposed C2D-Att for speaker verification which applies 2D convolution to obtain attention weights for both channel and frequency dimensions simultaneously. C2D-Att first pools time dimension by averaging, and then apply two consecutive 2D convolution modules to channel and frequency dimensions by introducing additional channel dimension which is increased to 8 and then back to 1. This results in channel-frequency attention weights which are multiplied to channel and frequency dimensions of preceding 2D convolution output. C2D-Att improves the speaker verification performance compared to fwSE by re-weighting channel and frequency dimensions simultaneously \cite{c2datt}. 

However, considering that channel dimension in CNN is permutable dimension where the convolution module's advantage capturing locality does not matter, it needs further verification if 2D convolution is the best option to apply channel and frequency attention on CNN. While CNN in C2D-Att applies 2D convolution kernel which finds local pattern across frequency and channel dimension, locality matters on frequency dimension only. Therefore, we experiment on joint application of attention on frequency and channel separately, without considering the locality of channel dimension using SE. To apply SE on two dimensions independently, we apply SE and tfwSE in series.


\section{Experimental Setups}
\label{sec:setup}
\subsection{Model Architecture}
The model architecture is based on CRNN model, composed of seven convolution layers followed by two bidirectional gated recurrent unit (GRU) then a FC layer. On the strong predictions, we apply class-wise median filter as post processing. In this work, the model using FDY conv replaced all 2D convolution except the first one. SE and C2D-Att modules are inserted after the activation and before the average pooling within the convolution blocks. They are applied on all convolution layers except the last layer in this work. It is because Hu \textit{et al.} has shown that SE module applies almost constant attention weights at the last layer thus it merely affects the model \cite{senet}.

\subsection{Implementation Details}
The overall implementation details follow the previous work \cite{FDY}, which could be referred on the official implementation code of which link is provided in the section \ref{sec:intro}. The experiments in this work are based on domestic environment sound event detection (DESED) dataset \cite{dcasebaseline}. DESED is composed of synthesized strongly labeled dataset, real weakly labeled dataset and real unlabeled dataset for training and validation. For test, real validation dataset, which is strongly labeled, is used. We do not use any external dataset. We trained each model with single NVIDIA RTX Titan GPU. For the results listed in this paper, the metrics are based on the best score among total 24 models from 12 separate training runs.

DESED is composed of 10 second audio data with 16 kHz sampling rate. We extract mel spectrogram as the input feature for SED model. The settings for mel spectrograms are as follows: 2048 points for number of fft, 256 points for hop length, Hamming window for windowing function, and 128 mel bins. 
Data augmentation methods applied are frame shift \cite{dcasebaseline}, mixup \cite{mixup}, time masking \cite{specaug} and FilterAugment \cite{filtaug}. Applying heavy data augmentation is crucial for training SED where real strongly labeled data is scarce \cite{mytechreport}. As we use three levels of datasets, strongly labeled/weakly labeled/unlabeled dataset, we apply mean teacher to leverage unlabeled dataset \cite{dcasebaseline, meanteacher}. 
We apply FilterAugment with different random parameters on student and teacher model to train SED model robust against FilterAugment.

\subsection{Evaluation Metrics}
Main evaluation metric employed in this study is the polyphonic sound detection score (PSDS) \cite{PSDS}, which considers the intersection between predictions and ground truth to decide if prediction is correct. PSDS also accounts for cross triggers induced by other sound events in the audio. PSDS utilizes area under curve (AUC) - receiver operating characteristic (ROC) curves, enabling comparison of sound event detection (SED) performances without the need for threshold optimization. In DCASE Challenge 2021, 2022 and 2023 Task 4, two variations of PSDS (PSDS1 and PSDS2) are utilized to evaluate SED systems \cite{dcasebaseline}. PSDS1 places emphasis on precise time localization by limiting tolerance for intersection criteria, while PSDS2 prioritizes accurate classification by penalizing cross triggers more. Additionally, we use collar-base F1 score (CB-F1) \cite{sedmetrics} for class-wise performance comparison, as PSDS cannot be obtained for single sound event. Both PSDS and CB-F1 are ranged between zero and one, and value closer to one indicates better SED performance.

\section{Results and Discussion}
\label{sec:result}
\subsection{Comparison of Attention Modules}
\begin{table}[t]
\caption{Performance and computational cost comparison between the baseline, frequency dynamic convolution and various frequency and channel attention methods on DESED real validation dataset.}
\centering
\setlength{\tabcolsep}{4.25pt}
\begin{tabular}{l|ll|lll}
\hline
\textbf{models}   & \textbf{params}      & \textbf{time}  & \textbf{PSDS1}           & \textbf{PSDS2}           & \textbf{CB-F1} \\
\hhline{=|==|===}
baseline          & \phantom{0}4.428M    & 3h 34m         & 0.409                    & 0.641                    & 0.520                    \\
+FDYconv          & 11.061M              & 6h 08m         & \textbf{0.446}           & \textbf{0.673}           & 0.525                    \\
\hline
+SE               & \phantom{0}4.537M    & 3h 49m         & \textbf{0.435}           & \textbf{0.654}           & 0.525                    \\
+tSE              & \phantom{0}4.537M    & 3h 52m         & 0.416                    & 0.643                    & 0.526                    \\
+fwSE             & \phantom{0}4.439M    & 3h 49m         & 0.411                    & 0.634                    & 0.522                    \\
+tfwSE            & \phantom{0}4.439M    & 3h 50m         & 0.415                    & 0.638                    & 0.509                    \\
\hline
+C2D-Att          & \phantom{0}4.429M    & 3h 53m         & 0.434                    & \textbf{0.659}           & 0.539                    \\
+tfwSE +SE        & \phantom{0}4.548M    & 4h 04m         & 0.437                    & 0.650                    & 0.532                    \\
+SE +tfwSE        & \phantom{0}4.548M    & 4h 06m         & \textbf{0.442}           & \textbf{0.657}           & 0.526                    \\
\hline
\end{tabular}
\label{tab:results}
\end{table}

\begin{figure*}[ht]
\centerline{\includegraphics[width=17.5cm]{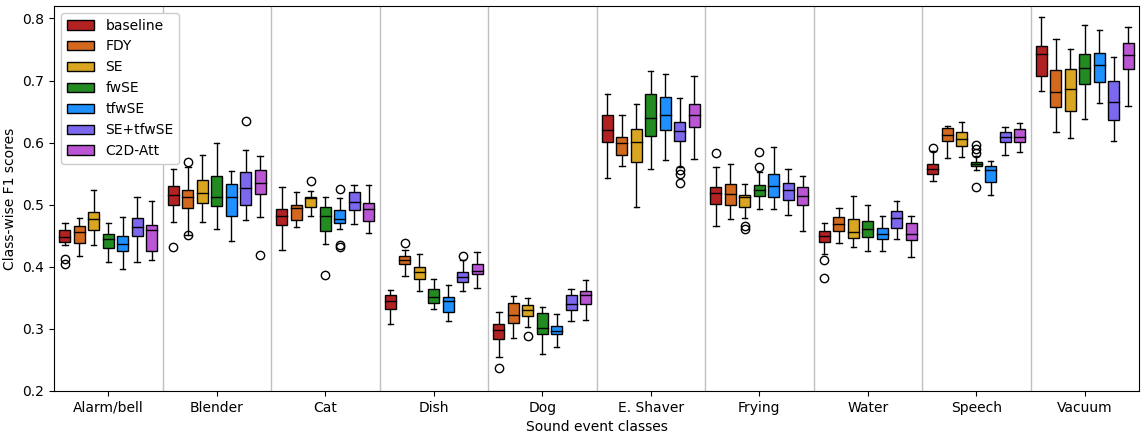}}
\caption{Box-plot of class-wise collar-base F1 scores by multiple models on DESED real validation dataset.}
\label{fig:cw-f1}
\end{figure*}

Table 1 shows performance and computational cost of SED models with various frequency-wise and channel-wise attention methods. Computational costs are described by the number of trainable parameters representing model size and training time representing computational efficiency. Note that we aim to achieve computational efficiency as close to the baseline as possible and much less than FDY conv. For comparison, SED model with FDY conv is listed in table \ref{tab:results} as well. Note that the results for FDY-CRNN differ from the results in previous paper due to minor changes in setting. When we compare the performance of SED model with SE variants, we can observe that conventional SE definitely outperforms the baseline. On the other hand, fwSE only slightly outperform the baseline for PSDS1 while their PSDS2 is worse than the baseline. Considering that SE is proposed to re-weight channel dimension and each channel is independent from each other while frequency depends on other frequency bins, re-weighting appears to be more effective on channel dimension than on frequency dimension. In addition, considering the parameter increase in the model, SE has increased model size significantly more thus it involved more computational resource to the model. While SE has increased model size by $\sim$2.5\%, fwSE has increased the model size by $\sim$0.25\%. Proposed tfwSE is slightly better than fwSE in terms of both PSDS1 and PSDS2. On the other hand, tSE only slightly outperforms the baseline and performs worse than SE. While re-weighting frequency dimension on each time frame has improved frame-wise sound event classification of SED, this effect seems to be not so significant. One explanation to this could be the effect of bi-GRU which processes time-varying information. Likewise, temporal dynamic convolution which applies time-adaptive kernel performed worse than FDY conv on SED \cite{tdycnn,FDY}. On the other hand, tSE failed to improve SE. Re-weighting a dimension separately on each time frame was not as effective on channel dimension. 

Results for methods applying attention simultaneously on channel and frequency dimensions, C2D-Att and joint applications of SE and tfwSE, are also listed in Table \ref{tab:results}. C2D-Att shows descent performance comparable to SE, with less parameters compared to SE. In addition, joint application of SE followed by tfwSE shows improvement over SE. While joint applicatoin of SE after tfwSE shows similar performance to SE, we could still conclude that application of attention methods simultaneously on channel and frequency dimensions are effective. Furthermore, the combination of SE and tfwSE achieves comparable results to FDY conv in terms of PSDS1, reaching 99.1\% of the PSDS1 by the model with FDY conv. Considering that high PSDS2 scores can be easily achieved using weakSED \cite{mytechreport}, we could regard that this model performs nearly as well as model with FDY conv. An interesting discovery is that while tfwSE degrades PSDS2 for the baseline model, the joint application of tfwSE after SE enhances PSDS2 compared to the model with SE alone. Moreover, considering that SE + tfwSE outperforms C2D-Att for PSDS1, 2D convolution considering locality of 2-dimensional patterns along frequency-channel dimensions is not as effective as separate consideration of channel and frequency dimensions. However, C2D-Att has advantage over SE + tfwSE in terms of the number of parameters which is increased by very small amount.

\subsection{Class-wise Performance Comparison}
In Fig. \ref{fig:cw-f1}, class-wise collar-based F1 scores on multiple models are shown as box-plot. Each box-plot is composed of class-wise F1 scores by 24 models from 12 separate training runs. Consistent to table \ref{tab:results}, SE performs better than fwSE and tfwSE on many classes in Fig. \ref{fig:cw-f1} as well. SE performed better than fwSE and tfwSE did on alarm/bell ringing, cat, dish, dog and speech while it performed worse on electric shaver, frying and vacuum cleaner. It seems that SE is stronger on transient and non-stationary sound events while it is weaker on quasi-stationary sound events, similar to FDY conv \cite{FDY}. That is to say, while fwSE and tfwSE re-weight frequency dimension to address frequency dimension, they are stronger on quasi-stationary sound events than on non-stationary sound events. SE + tfwSE shows similar tendency with SE, but slightly better performance in general. Thus SE + tfwSE perform relatively better on non-stationary sound events and relatively worse on quasi-stationary sound events as well. C2D-Att also shows similar tendency with SE but it shows better performance on electric shaver and vacuum cleaner. Note that PSDS is an intersection-based score while the box-plots are based on collar-based score, there are slight discrepancy between table \ref{tab:results} and Fig. \ref{fig:cw-f1}.


\section{Conclusion}
\label{sec:conclusion}
In conclusion, we experimented on various frequency and channel attention methods to enhance SED performance while minimizing computational cost. The study addressed the challenge of effectively addressing the frequency dimension of audio data by leveraging attention methods. The attention methods demonstrated comparable performance to the previous approach of FDY conv, while reducing the computational cost and improving practicality. In addition, we performed class-wise performance of the attention methods to further analyze the characteristics of SED models with different attention methods. Future research could aim to optimize the proposed attention methods by applying them jointly with FDY conv either to push the performance even more or to find balance between computational cost and the performance.

\section{Acknowledgment}
\label{sec:acknowledgment}
We would like to thank Junhyeok Lee from Supertone Inc. for valuable discussions.

\bibliographystyle{IEEEtran}
\bibliography{refs}

\end{sloppy}
\end{document}